\documentclass[prb,twocolumn,amsmath,amssymb,amsfonts,a4paper]{revtex4}

\newcommand{\be}{\begin{equation}}
\newcommand{\ee}{\end{equation}}

\usepackage{graphicx,amsmath}
\usepackage{dcolumn}

\newcommand{\ltsim}{\protect\raisebox{-0.5ex}{$\:\stackrel{\textstyle <}
        {\sim}\:$}}

\begin{document}

\title{Precision shooting: Sampling long transition pathways}

\author{Michael Gr\"unwald}
\affiliation{Faculty of Physics and Center for Computational Materials Science, University of Vienna, Boltzmanngasse 5, 1090 Vienna, Austria}
\author{Phillip L.~Geissler}
\affiliation{Department of Chemistry, University of California at Berkeley, Berkeley, California 94720}
\affiliation{Chemical Sciences Division, Lawrence Berkeley National
Lab, Berkeley, California 94720} 
\author{Christoph Dellago}
\affiliation{Faculty of Physics and Center for Computational Materials Science, University of Vienna, Boltzmanngasse 5, 1090 Vienna, Austria}

\begin{abstract}
The kinetics of collective rearrangements in solution, such as protein
folding and nanocrystal phase transitions, often involve free energy
barriers that are both long and rough. Applying methods of transition
path sampling to harvest simulated trajectories that exemplify such
processes is typically made difficult by a very low acceptance rate
for newly generated trajectories.
We address this problem by introducing a new 
generation
algorithm based on the linear short-time
behavior of small disturbances in phase space. Using 
this ``precision shooting'' technique,
arbitrarily small disturbances can be 
propagated in time,
and any desired acceptance ratio of shooting moves can be obtained. We
demonstrate the method 
for a simple but computationally problematic isomerization
process in a dense liquid
of soft 
spheres. We
also discuss its applicability to barrier
crossing events involving metastable intermediate states.
\end{abstract}

\maketitle

% ************************************************************************************

\section{Introduction}
Transition path sampling (TPS) is a versatile and efficient set of
computational techniques for the study of rare
events.\cite{Dellago1998,Dellago2002,Dellago2006,Dellago2008} It has been
successfully used to 
reveal the microscopic
mechanisms of processes as diverse
as autoionization in liquid water \cite{Dellago2001}, structural
transformations in nanocrystalline solids \cite{Grunwald2007}, and
folding of small proteins.\cite{Bolhuis2006} 
The purpose of this paper is to propose a new shooting algorithm which
can greatly increase the efficiency of TPS when transit times of
activated trajectories greatly exceed the picosecond time scale of
phase space stability.

At its core TPS is a Monte Carlo procedure enabling
a random walk in the ensemble of 
pathways 
that cross
a free energy barrier
between two metastable states (denoted A and B). While this 
sampling
is strongly biased towards reactive trajectories, it leaves the
underlying dynamics of the system unchanged. Thus, the result of a TPS
simulation is a representative set of true dynamical pathways,
weighted as if they were excerpted from an extremely long, unbiased
simulation of equilibrium dynamics. Many analytical tools
have been developed to extract from such a collection of trajectories
useful molecular
information about the process of interest.\cite{Dellago2008}

The algorithm 
typically
used to construct such a random walk is
called \emph{shooting} \cite{Dellago2002}. Here, a point along a given
reactive trajectory is randomly selected and slightly changed; for
instance, one might change the velocities of all particles by a small
random
number drawn from 
a symmetric
distribution. Using the dynamical rules of the system, this \emph{shooting point} is then propagated
forward and backward in time to obtain a complete new trajectory. If
this new trajectory still connects A with B it is accepted and used as
a basis for the next shooting move; otherwise it is rejected.
 
The efficiency of this algorithm in 
exploring
the transition path ensemble is based on 
a
balance between the
intrinsic
instability of 
complex
dynamical systems and the local character of the
shooting move: Small disturbances grow exponentially 
quickly in time, leading 
to separation of trajectories typically 
within
a few picoseconds. Nonetheless, if the 
disturbance is small, the
new trajectory will be locally similar to the old one and 
is therefore likely to surmount the barrier between A and B;
such shooting moves will be
accepted frequently. Just 
as with conventional
Monte Carlo moves in
configuration space, maximum efficiency can often be obtained by
adjusting the size of the disturbance 
to achieve an acceptance probability of roughly 40\%.
\cite{Dellago2002}

Shooting moves are best suited for the study of systems that relax
quickly (within the picosecond time scale of trajectory separation)
into their product state after reaching the top of the barrier.
Many interesting processes, like the nucleation of first order
phase transitions 
or conformational change 
in complex molecules,
proceed much more slowly from the transition state.
In TPS simulations of such systems, shooting moves
must be made extraordinarily subtle in order to stand a reasonable
chance of connecting reactant and product states. As a matter of
practice, however, disturbances cannot be made arbitrarily small
due to the limited machine precision of floating point numbers.
Lacking an ability to control the degree of global separation between
trajectories, TPS methods are severely compromised in efficacy.
The demonstrated computational advantages of importance sampling
in trajectory space lose appeal when offset by the wasted effort
of generating a vast excess of non-reactive paths.

In a recent paper\cite{Bolhuis2003}, Bolhuis addressed this problem by
modifying slightly the rules that propagate a system in
time. Specifically, a weak stochastic component was added to the
dynamics, removing the unique correspondence between a trajectory's
past and its future. It thus became possible to resample
only parts of an existing pathway, leading to much higher
acceptance probabilities for shooting moves and, Bolhuis reports,
significant improvement in sampling efficiency.\cite{Bolhuis2003}

In this paper, we show that it is possible to perform productive
shooting moves for arbitrarily long transition paths without modifying
a system's natural dynamics. Our technique for introducing and
propagating extraordinarily small disturbances is based on the simple
dynamics of small perturbations in phase space. We explain the method
and offer a straightforward algorithm for implementation in
Sec.~\ref{sec:method}. Use of the technique is demonstrated in
Sec.~\ref{sec:isom} for a simple isomerization process in a dense
liquid that, by construction, involves diffusive dynamics on a rugged
barrier. In Sec.~\ref{sec:meta} we examine limitations of the method
by considering reactive dynamics that pass through highly metastable,
obligatory intermediate states.

% ************************************************************************************

\section{\label{sec:method}Linearized dynamics of small perturbations}
 
\subsection{Exponential divergence of trajectories}
\begin{figure}
\includegraphics[width=0.4\textwidth]{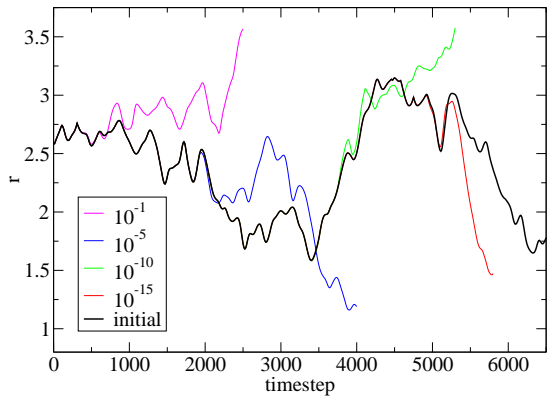}
\caption{\label{diverge_fig} Distance $r$, in reduced units, 
between
two particles in a liquid of 108 soft spheres\cite{testsys} as a
function of time 
along
a reference trajectory and four shooting trajectories. At time zero,
displacements of 
various
size are added to 
particle velocities
in the reference trajectory. Because 
phase space disturbances grow exponentially in time,
decreasing the shooting displacement by 
successive
orders of magnitude 
results in 
only
linear increase in 
the time that elapses before trajectories separate.
Shooting displacements smaller than $10^{-15}$ can not be resolved in
double precision; the resulting shooting trajectories will exactly
retrace the initial one.}
\end{figure}
In a TPS simulation of a system evolving with deterministic dynamics,
a trajectory $X$ of length $\tau$ consists of a number of ``snapshots''
$x_{i\Delta t}$, which are separated by a time step $\Delta t$,
\begin{equation}
X = \lbrace x_0,x_{\Delta t},x_{2\Delta t},\dots,x_{\tau} \rbrace\;.
\end{equation}
Here, the time slices $x_{i\Delta t}$ are full phase space vectors,
detailing
the positions and velocities of all
particles. Subsequent time slices are related by
\begin{equation}
x_{(i+1)\Delta t} = \Phi(x_{i\Delta t})\;,
\end{equation}
where the function $\Phi$ propagates the system for one time step.

Consider now a shooting move, in which a small disturbance $\delta
x_0$ is added to the shooting 
point $x_0$ to obtain state $y_0 = x_0 + \delta x_0$ of the shooting trajectory $Y$. (To simplify notation, we will
assume the shooting point to be $x_0$, the initial state of the
trajectory, throughout 
this section. The algorithm we will describe applies transparently to
shooting points at any chosen time along the trajectory.)
Usually the perturbation $\delta
x_0$ affects only momentum space, but changing the positions of the
particles can be useful in some cases \cite{Dellago2002}. The
perturbed point $y_0$ is then propagated for a number of $n$ time
steps to obtain $y_{t} = \Phi_{t}(y_0)$, where $t = n\Delta t$ and
$\Phi_t$ refers to the $n$-fold application of the time step
propagator. We define the time-evolved disturbance $\delta x_t$ by
subtracting the old trajectory from the new one, $\delta x_t =
y_t-x_t$. Due to the dynamic instability of the system, perturbations
grow exponentially 
in time,
\begin{equation}\label{exp_eq}
 |\delta x_t| \approx |\delta x_0|\, e^{\lambda t}\;.
\end{equation}
Here, $\lambda$ is the largest Lyapunov exponent of the
system.\cite{Posch1989} For typical fluid systems, $1/\lambda\approx
1\,\mathrm{ps}$. 

\begin{figure}
\includegraphics[width=0.4\textwidth]{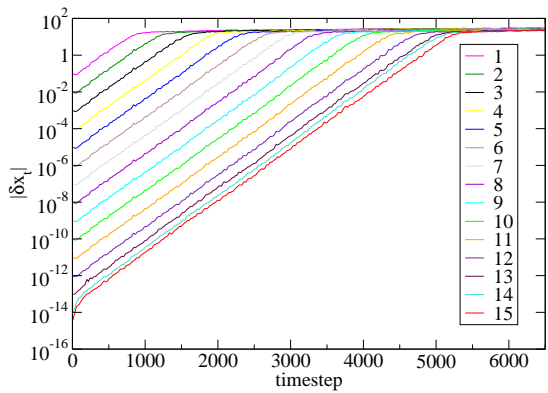}
\caption{\label{expgrow_fig} Time evolution of small displacements
$\delta x_t$ in a liquid of 108 soft spheres\cite{testsys} for
disturbances of various sizes, $\vert\delta x_0\vert=10^{-\alpha}$ (legend values indicate values of $\alpha$).
All displacements
grow exponentially with the same rate,
up to the 
time 
where trajectories separate. The maximum possible value of $\vert\delta x_t\vert$ is determined by the dimensions of the simulation box. Note that adjacent lines are equidistant in the linear regime,
except for displacement sizes smaller than $10^{-12}$.
Although these smallest displacements yield trajectories that can in
practice be distinguished from the base trajectory, limited numerical
precision introduces rounding errors that degrade computational
estimates of linear divergence.}
\end{figure}

We wish to 
control 
precisely
the time it takes for a small perturbation to reach a size of order 1,
at which point
the new trajectory will be essentially separated from the old
one.
This time determines the 
probability that the new trajectory will be reactive and therefore
acceptable.
Because of 
subsequent
exponential growth, $|\delta
x_0|$ must be decreased by many orders of magnitude to increase the
separation time of trajectories by 
even
a few picoseconds (see Figs.~\ref{diverge_fig} and
\ref{expgrow_fig}). 
With the standard double precision format for
representing floating point numbers on a computer,
however, the smallest number that can be added to 1.0 to give a result
distinguishable
from 1.0 is of the order of $10^{-15}$, and numerical
results become unreliable at values of $|\delta x_0|$ well above this
limit. 
(We assume throughout this paper that a system of units has been
chosen such that typical numerical values of coordinates and momenta
are of order 1.)
Especially when the total length of the transition path is
significantly longer than a few picoseconds, the limited range of
practical displacement sizes constitutes a severe sampling problem:
Shooting moves will only be accepted from points in the vicinity of
the barrier top; otherwise, new trajectories will simply return to the
stable state they came from and be rejected. As the system may stay
near the \emph{a priori} unknown barrier top only for a small fraction
of the total transition time, sampling can break down completely. In
these cases, 
implementing shooting
displacements of arbitrarily small size would be very helpful.

\subsection{Dynamics in the linear regime}
\begin{figure}
\includegraphics[width=0.4\textwidth]{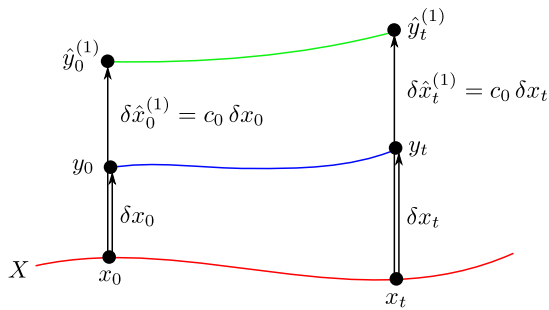}
\caption{\label{shoot_fig} 
Time evolution of shooting moves from a point $x_0$ on the reference
trajectory $X$.  The two shooting displacements $\delta x_0$ and
$\delta\hat x_0^{(1)}$, as vectors in high-dimensional phase space, point in the
same direction but have different magnitudes, $\delta\hat x_0^{(1)} = c_0\,\delta
x_0$.  At a later time $t$ short enough that first-order perturbation
theory remains valid, these displacements remain proportional, $\delta
x_t^{(1)} = c_0\,\delta x_t$.  The displacement of interest $\delta x_t$, no
matter how small, can thus be constructed in the linear regime
simply by dividing $\delta\hat x_t^{(1)}$  by the original scaling factor $c_0$.}
\end{figure}
We propose to solve this problem by using perturbation theory to
follow the time evolution of the {\em displacement} vector $\delta x_0$
itself, up to the point where it 
grows large enough to allow accurate evaluation of the sum $x_t + \delta x_t$.
Expanding $y_t$ around $x_0$, we 
obtain
\begin{eqnarray}
\delta x_t = y_t-x_t &=& \Phi_t(x_0+\delta x_0)-\Phi_t(x_0)=\nonumber\\
&=&\frac{\partial\Phi_t(x_0)}{\partial x_0}\,\delta x_0 + \mathcal{O}\left[(\delta x_0)^2\right]\;.
\end{eqnarray}
For small displacements $|\delta x_0|<10^{-15}$, the linear
approximation 
is for all practical purposes exact on the scale of a single time step,
\begin{equation}\label{lineq}
\delta x_t=S\,\delta x_0\;, 
\end{equation}
where the matrix $S$ is given by
\begin{equation}
S = \frac{\partial\Phi_t(x_0)}{\partial x_0}\;.
\end{equation}
To integrate $\delta x_0$ forward in time according
to equation (\ref{lineq}), the equations of motion for the matrix $S$
could in principle be solved numerically.\cite{Dellago1996} Doing so in practice would
be cumbersome, requiring calculation of all second derivatives of the
interaction potential with respect to particle positions.
We propose a much simpler approach for advancing $\delta x_t$, inspired
by methods for computing Lyapunov exponents in systems whose interaction
potentials lack well-defined second derivatives. Our implementation
is illustrated in Fig.~\ref{shoot_fig}.

Instead of integrating the small perturbation $\delta x_0$, we follow
the time evolution of a 
related
perturbation $\delta\hat x_0^{(1)}$, which 
{\em is} 
large enough to be added at the shooting point and 
propagated
in the usual way, 
i.e., by integrating Newton's equation of motion for $\hat y_0^{(1)} = x_0 +
\delta \hat x_0^{(1)}$. We use a superscript for $\hat y^{(i)}$ and $\delta \hat x^{(i)}$ because in the following we will consider a family of different perturbed trajectories $\hat Y^{(i)}=\lbrace \hat y_0^{(i)},\dots,\hat y_\tau^{(i)}\rbrace$, with $\hat y_t^{(i)}=x_t+\delta\hat x_t^{(i)}$.  Exploiting the linearity described by
Eq.~(\ref{lineq}), we choose $\delta\hat x_0^{(1)}$ to be in the same
direction (in the high-dimensional phase space) as $\delta x_0$,
\begin{equation}
 \delta\hat x_0^{(1)} = c_0\,\delta x_0\;,
\end{equation}
where $c_1$ is a 
scalar constant.
If $\delta\hat x_0^{(1)}$ is 
also
small enough 
to justify the linear approximation of Eq.~(\ref{lineq}), 
\begin{equation}
 \delta\hat x_t^{(1)} = S\,\delta\hat x_0^{(1)}\;,
\label{equ:lineqy}
\end{equation}
then
the initial relationship between $\delta x_0$ and $\delta\hat x_0^{(1)}$
holds also at a later time $t$,
\begin{equation}\label{lin_eq}
\delta x_t = S\,\delta x_0 = \frac{1}{c_0}\,S\,\delta\hat x_0^{(1)} = \frac{1}{c_0}\,\delta\hat x_t^{(1)}\;. 
\end{equation}
In the linear regime it is thus possible to follow the time evolution
of arbitrarily small displacements $\delta x_0$ by monitoring larger,
proportional displacements.

The linear approximation for the ``helper'' displacement $\delta\hat x_t^{(1)}$
in Eq.~(\ref{equ:lineqy}) will of course remain valid for only a short
time $t_{\rm lin}^{(1)}$, typically less than 1 ps. Our interest in
the trajectory $\hat Y^{(1)}$, however, is only as a proxy for the evolution of
smaller displacements that cannot be represented explicitly. As
$\delta\hat x_t^{(1)}$ approaches the boundary of the linear regime, $t \ltsim
t_{\rm lin}^{(1)}$, we may therefore switch our attention to a
different helper trajectory $\hat Y^{(2)}$, one whose displacement is initially
too small to be of practical use but by the time $t_{\rm lin}^{(1)}$
grows large enough to be represented explicitly. The new displacement
$\delta\hat x_t^{(2)} = \hat y_t^{(2)}-x_t$ can be obtained at any time $t < t_{\rm
lin}^{(1)}$ simply by scaling $\delta\hat x_t^{(1)}$ appropriately, $\delta\hat x_t^{(2)}
= \delta\hat x_t^{(1)}/c_1$. Because it is initially smaller than $\delta\hat x_t^{(1)}$,
it will remain in the linear regime for a longer time, $t_{\rm
lin}^{(2)}>t_{\rm lin}^{(1)}$. For times $t > t_{\rm lin}^{(1)}$ we
therefore proceed by integrating standard equations of motion for $\hat Y^{(2)}$
until it approaches the boundary of the linear approximation. At that
point we repeat the procedure, scaling back the displacement to switch
attention to yet another helper trajectory. 

In effect we monitor a single displacement from the reference
trajectory whose magnitude is periodically scaled down such that the
linear approximation is always valid. In this way we can monitor the
time evolution of an arbitrarily small disturbance. The corresponding
trajectory will be numerically indistinguishable from the reference
trajectory as long as the displacement's magnitude is smaller than
$\sim 10^{-15}$. At later times the displaced trajectory can be
distinguished, and its dynamics can be safely computed by integrating
equations of motion in the usual way.

\subsection{Algorithm}
\begin{figure}
\includegraphics[width=0.4\textwidth]{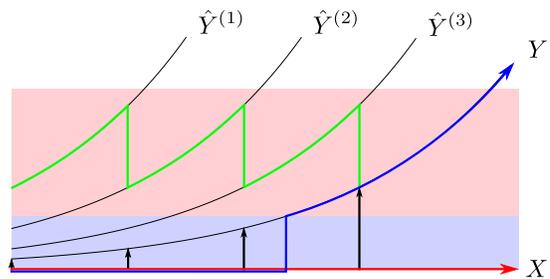}
\caption{\label{algo_fig} Precision shooting algorithm 
for generating a trial trajectory $Y$ (blue curve) whose initial
displacement from the base trajectory $X$ (red line) is extraordinarily
small. Points in the light blue field (whose extent is $\approx
10^{-15}$) cannot be numerically distinguished from the reference
trajectory. Until the trial trajectory exits this region, its time
evolution is calculated by proxy using ``helper'' trajectories $\hat Y^{(i)}$ (thin black curves).  Displacements of trial (black arrows) and helper trajectories
from the base trajectory are related by proportionality as long as
they remain within the linear regime, represented by the light red
field. To preserve this simple relationship, the displacement of the
helper trajectory $\hat Y^{(i)}$ from the reference trajectory is scaled back
when it threatens to leave the region of linear dynamics, effectively switching the system to the next helper trajectory $\hat Y^{(i+1)}$. By following the systems dynamics along many sections of helper trajectories (green curve) and by keeping track of the rescaling factors, one can accurately construct the state of the trial trajectory once it
becomes distinguishable from the reference trajectory.
The result of this shooting move is a trial trajectory $Y$
that is numerically identical to the base trajectory $X$ over a certain
length of time and then emerges from it in the correct way.}
\end{figure}
This insight suggests the following algorithm, which implements a
shooting trajectory $Y$, whose initial deviation $\delta x_0$ from the
base-trajectory $X$ is smaller than the precision limit. This is done
by monitoring ``helper'' trajectories $\hat Y^{(j)}$, which are obtained by repeated rescaling.
For an illustration of this algorithm see
Fig.~\ref{algo_fig}.
\begin{enumerate}
\item At the shooting point $x_0$, add a displacement $\delta\hat x_0^{(1)}$ of
fixed size $|\delta\hat x_0^{(1)}|=\sigma$. The displacement $\delta\hat x_0^{(1)}$ is parallel to $\delta x_0$ and larger by a factor of $c_0$.
\item 
Propagate
the point $\hat y_0^{(1)} = x_0+\delta\hat x_0^{(1)}$ forward 
in time
for $n$
time steps, 
corresponding
to a time interval $t=n\Delta t$.
\item 
Compute the factor $c_1 =|\delta\hat x_t^{(1)}|/|\delta\hat x_0^{(1)}|=|\hat y_t^{(1)}-x_t|/|\delta\hat x_0^{(1)}|$
quantifying the divergence from the reference trajectory. Switch to a new helper trajectory by setting $\hat y_t^{(2)} = x_t+\delta\hat x_t^{(1)}/c_1= x_t+\delta\hat x_t^{(2)}$. Store the factor $c_1$.
\item
Propagate the new displacement forward in time by integrating the
equations of motion for $n$ steps beginning from $\hat y_t^{(2)}$.  Calculate
and store the factor $c_2 = |\delta\hat x_{2t}^{(2)}|/|\delta\hat x_t^{(2)}|$.  

\item Iterate step 4, each time beginning from $\hat y_{(j-1)t}^{(j)}$ and
rescaling by the factor $c_j = |\delta\hat x_{jt}^{(j)}|/|\delta\hat
x_{(j-1)t}^{(j)}|$. At every iteration compute
the displacement of interest, $\delta x_{jt}=C_j^{-1}\,\delta\hat x_{jt}^{(j)}$, where 
$C_j = \Pi_{k=0}^{j-1}c_k$
is the product of all factors 
used for rescaling so far. To store the current point along the actual
shooting trajectory, compute $y_{jt} = x_{jt} + \delta x_{jt}$. As long as
$|\delta x_{jt}| \lesssim 10^{-15}$, $y_{jt}$ will be numerically identical
to $x_{jt}$.
\item If $|\delta x_{jt}| > \sigma$, the 
actual shooting
displacement is large enough to
be treated in the usual way: Set $y_{jt} = x_{jt} + \delta x_{jt}$ and
integrate 
equations of motion from
this point without further rescalings (ceasing iteration of step 4). 
\end{enumerate}
Note that for shooting moves conducted at points other than $x_0$, the
procedure must be repeated backward in time to obtain a complete
shooting trajectory. In the following we discuss the accuracy of this
scheme and give recommendations for 
choosing values
of $\sigma$ and $n$.

\subsection{Validity of the linear approximation}
\begin{figure}
\includegraphics[width=0.4\textwidth]{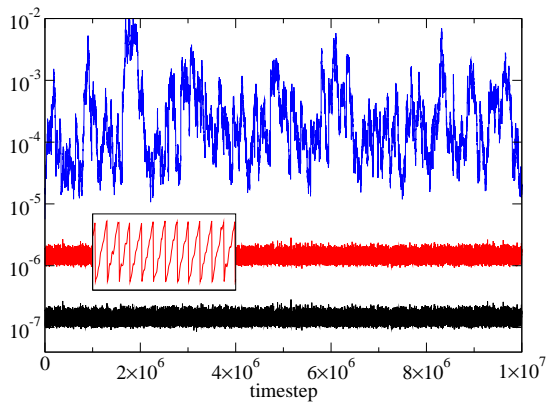}
\caption{\label{linreg_fig} Time evolution of the magnitude of two shooting
displacements for a fluid of WCA particles.\cite{testsys}
Displacements of size $|\delta\hat x_0^{(1)}|=10^{-7}$ (black) and
$|\delta\hat x_0^{(2)}|=10^{-6}$ (red) are initially proportional (pointing in the same direction in phase
space). Both are rescaled to their initial size every 100 time steps (see inset for a 
magnified view) to preserve this linear relationship. Deviation from proportionality is quantified by
the relative error $\epsilon(t)$ (blue) defined in Eq.~(\ref{error_eq}).}
\end{figure}
The above algorithm is exact only if the linear approximation of Eq.~(\ref{lineq}) holds. For perturbations of finite size, deviations from this approximation occur. The question thus arises, how accurate an approximation is this approach for
propagating small disturbances? More specifically, to what extent do
helper displacements remain proportional to the actual shooting
displacements of interest? One could certainly imagine that the fast growth of small non-linearities rapidly erodes the linear relationship on which we depend. Here we present evidence from computer simulations that
proportionality of small displacements can hold in practice over very long time scales.

Figure \ref{linreg_fig} shows the time evolution of two 
proportional disturbances. The initial displacement
vectors point in the same direction of phase space but have
different magnitudes, $|\delta\hat x_0^{(1)}|=10^{-7}$ and
$|\delta\hat x_0^{(2)}|=10^{-6}$.
The respective shooting trajectories 
were propagated
independently,
and the displacements from the base trajectory 
were
rescaled to their initial length every 100 time steps. 
For a perfectly linear time evolution, these displacements remain proportional at
all later times. In practice the vectors
$\delta\hat x_t^{(1)}$ and $\delta\hat x_t^{(2)}$ will develop a nonzero angle due to non-linearities. To quantify this deviation
from parallel alignment, we define
\begin{equation}\label{error_eq}
\epsilon(t)=\frac{|\delta\hat x_t^{(1)} - \delta\hat x_t^{(2)}/10|}{|\delta\hat x_t^{(1)}|}\,.
\end{equation}
As shown in Fig.~\ref{linreg_fig}, the relative error $\epsilon(t)$
does not grow above a low level even for very long simulation runs. 

This long time stability of aligned disturbances holds over a broad
range of displacement sizes between $10^{-10}$ and $10^{-3}$.
Values of $\epsilon(t)$ can be somewhat larger than for the specific
case plotted in Fig.~\ref{linreg_fig} but on average do not grow larger
than $10^{-3}$ for any case.
The error is insensitive to the choice of the rescaling interval
$n\Delta t$, as long as the displacements stay smaller than approximately
$10^{-2}$, where the linear regime breaks down. For displacements
smaller than $10^{-10}$, rounding errors become 
problematic, and phase space displacements do not remain parallel to a
good approximation. One might expect the region of long time stability to extend to even lower levels, closer to the precision limit of $10^{-15}$. However, the total achievable accuracy in a computer simulation depends, among other
factors, on the details of the integrator $\Phi$, the size of the time
step, and the dimensionality of the system, and can lie well above the
precision limit of $10^{-15}$. We find that for the particular system
used here, the precision level is about $10^{-12}$.  

In the light of these observations, a value of $\sigma=10^{-6}$ as initial magnitude for helper displacements seems appropriate. This choice lies midway between the upper limit of the linear regime (approximately $10^{-2}$), and the point where rounding errors become dominant (approximately $10^{-10}$). As the accuracy of the rescaling scheme is quite insensitive to the frequency of rescalings, many equally good choices of $n$ are possible. As a starting point, a value of $n$ which leads to rescalings every time the displacements have doubled their size is advisable.

\begin{figure}
\includegraphics[width=0.4\textwidth]{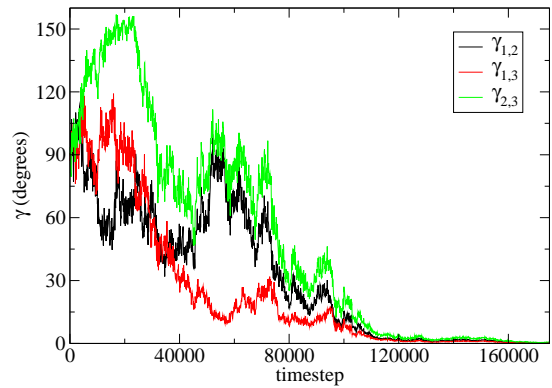}
\caption{\label{angle_fig} 
Relative orientations of three phase space displacement vectors
($\delta\hat x_t^{(1)}$, $\delta\hat x_t^{(2)}$, and $\delta\hat x_t^{(3)}$) for a WCA
fluid.\cite{testsys} We plot the angle $\gamma_{i,j} =
\cos^{-1}(\delta\hat x_t^{(i)} \cdot \delta\hat x_t^{(j)}/|\delta\hat x_t^{(i)}|
|\delta\hat x_t^{(j)}|)$ between each pair of displacements as a function
of time. Displacements differ in
initial direction and size ($|\delta\hat x_0^{(1)}|=10^{-8}$, $|\delta\hat
x_0^{(2)}|=10^{-8}$, $|\delta\hat x_0^{(3)}|=10^{-7}$) and are rescaled to their
original sizes at different intervals ($n_x=100$, $n_y=200$,
$n_z=500$).}
\end{figure}

The fact that $\epsilon(t)$ does not show any  
systematic long-time growth in Fig.~\ref{linreg_fig}
seems surprising. After all, no constraint is 
imposed
on the direction of the displacement vectors.
Why does an 
accumulation of errors 
not
eventually lead to 
decoupling and
$\epsilon(t)\approx 1$?
Stability of the 
precision shooting
algorithm is 
in fact a simple and
direct consequence of the collective
dynamics of displacements in the linear regime. In Figure
\ref{angle_fig} we plot the angles between three periodically rescaled
shooting displacement
vectors
of different size and random initial
direction. Eventually, they all 
rotate
into the same direction, which is
associated with the 
largest Lyapunov exponent $\lambda$ of
the system.
The time scale on which the directions of different
displacement vectors converge is on the order of $1/\Delta\lambda$,
where $\Delta\lambda$ is the difference between
the first and second 
largest
Lyapunov exponents.\cite{Dellago2002a} It is
because of this convergence, that the difference vector $\delta\hat
x_t^{(1)}-\delta\hat x_t^{(2)}/c$ between two 
proportional
displacements with initially
identical direction will stay small.

We point out that this property constitutes a main difference of our
method over the stochastic scheme introduced by Bolhuis\cite{Bolhuis2003} and similar algorithms. Consider, for
instance, the following simple algorithm that can be viewed as a
smooth version of the stochastic scheme by Bolhuis:
\begin{itemize}
\item Choose a shooting point $x_\mathrm{s\Delta t}$.
\item A fixed number of timesteps $n$ earlier and later, at the points $x_{(s+n)\Delta t}$ and $x_{(s-n)\Delta t}$, add a displacement of $10^{-15}$ to one velocity component of one particle.
\item Integrate the points $x_{(s+n)\Delta t}$ and $x_{(s-n)\Delta t}$ forward and backward in time, respectively, to get a complete new trajectory.
\end{itemize}
Just like the precision shooting algorithm, this simple scheme results
in a shooting trajectory that is numerically identical to its base
trajectory for a certain period of time. However, the emerging
separation between base and shooting trajectories will not be consistent
with a shooting move conducted at $x_0$, but rather with two 
uncorrelated
shooting moves at 
$x_{(s+n)\Delta t}$ and $x_{(s-n)\Delta t}$. 
Our algorithm, on the
other hand, correctly reproduces the 
correlated forward and backward
dynamics of a displacement
introduced at 
$x_{s\Delta t}$.

\section{\label{sec:isom}A simple test system}
\label{app_sec}
We demonstrate the precision shooting algorithm on a simple
isomerization process of a 
solvated diatomic molecule in three dimensions.

\begin{figure}
\includegraphics[width=0.4\textwidth]{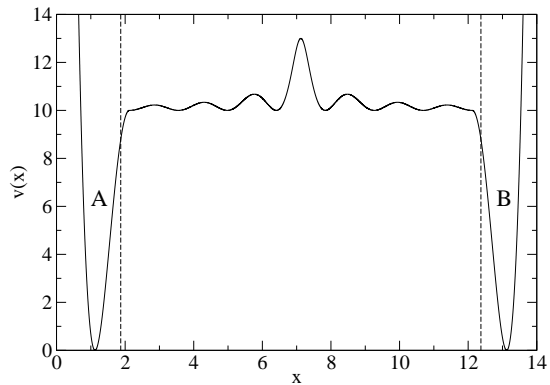}
\caption{\label{pot_fig} 
Potential energy $v(x)$ of interaction between particles \#1 and \#2, comprising the diatomic
molecule in our model isomerization process, plotted
as a function of the difference $x=x_2-x_1$ between 
their $x$-coordinates. The dashed lines mark the boundaries of the minima A and B.}
\end{figure}

Our test system consists of 389 particles interacting \emph{via} the
WCA\cite{Weeks1971} potential. We use conventional reduced units, with particle mass and potential parameters $\sigma$ and $\epsilon$ all set to unity. Particles 
\#1 and \#2 do not interact \emph{via} the WCA potential, but
are
bonded
through a one-dimensional potential with two deep minima
separated by a rough barrier (see Fig.~\ref{pot_fig}):
\begin{equation}
v(x)=\left\{ \begin{array}{ll} 	h_1\left[1-q(x)^2/w^2\right]^2 & \quad\mathrm{if}\; q(x)<0\,,\\ 
				h_1\left[1-(q(x)-b)^2/w^2\right]^2 & \quad\mathrm{if}\; q(x)>b \,,\\
				h_1 + h_2\frac{\cos^2\left[a(q(x)-b/2)\right]}{\sqrt{1+ga^2(q(x)-b/2)^2}} & \quad\mathrm{else}\,.
\end{array}\right. 
\end{equation}
Here,  $q(x) = x-(r_c+w)$, $x=x_2-x_1$ is the difference between the $x$-component of the position of the bonded particles, $r_c=2^{1/6}$ is the cutoff of the WCA potential, $w=1$ determines the width of the minima, $b=10$ and $h_1=10$ are the length and height of the barrier in between, respectively, and the constants $h_2=3$, $a=7\pi/b$, and $g=2$ determine the shape of the barrier. The potential and its derivative are continuous by construction.

To 
speed computation,
we 
borrow
a trick 
from Bolhuis'
work:
\cite{Bolhuis2003} Particle \#2 is considered to 
lie always to the right
of particle \#1, hence $x>0$. This
choice,
together with
the one-dimensionality of $v(x)$, allows us to choose a simulation box with dimensions $14.4\times6\times6$. The resulting
particle density is 0.75, the total energy per particle is 1.0, and the temperature is 0.45, as gauged by average kinetic energy. We use the velocity Verlet algorithm\cite{Frenkel2002} to integrate the equations of motion with a time step of 0.002.
\begin{figure}
\includegraphics[width=0.4\textwidth]{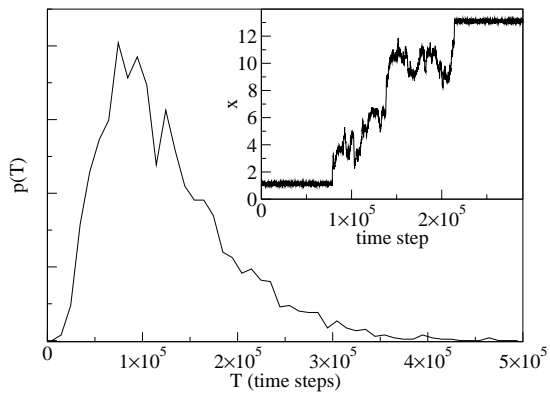}
\caption{\label{dist_fig} Distribution of transition times $T$ 
for our model isomerization process, as gauged from 2500
trajectories initiated at the barrier top.
Inset: Difference $x=x_2-x_1$ between 
the $x$-coordinates of particles \#1 and \#2 as a function of time for a typical trajectory.}
\end{figure}

We are interested in sampling the transition of the dimer from the
``contracted''
minimum A at $x_\mathrm{A}=r_\mathrm{c}$ to the 
``extended''
minimum B at $x_\mathrm{B}=r_\mathrm{c}+b+2w$. The dimer is defined to
be in state A for $x<x_\mathrm{A}+0.75w$ and in state B for
$x>x_\mathrm{B}-0.75w$ (see Fig.~\ref{pot_fig}). 
Because the system is dense, and the barrier is both long and rough,
relaxation from the transition state into either stable minimum is
quite protracted. 

In conducting TPS simulations it is important that sampled
trajectories are not shorter than typical spontaneous barrier-crossing
events.\cite{Dellago2002} We determine this typical
duration for our simple model system by initiating many
straightforward molecular dynamics simulations with the dimer bond
length set at $x=r_\mathrm{c}+w+b/2$, corresponding to the middle of the
barrier. Integrating the equations of motion forward and backward in time
yields a representative sample of the transition path ensemble. For a particular trajectory, the transition time $T$ is the time the system spends between regions A and B. The
resulting
distribution of transition
times 
is
plotted in Fig.~\ref{dist_fig}. For 
TPS simulations, we
choose a total trajectory length of $3\times10^5$ time steps, long
enough to 
include 98\% of the natural transition path ensemble. The bias
of our sampling to short transitions is therefore minor.

Although the artificial potential energy landscape studied here does
not directly represent any physical system of interest, it
nevertheless shares with many real systems features that lead to long
transition pathways and make straightforward application of TPS
methods ineffective. In our view roughness of the barrier region is an
important ingredient. Models featuring long but flat barriers, such as
that of Ref.~\cite{Bolhuis2003}, should not in fact pose any severe
problems for path sampling via the standard shooting move. Assuming
that motion atop such a flat barrier is diffusive in nature, and that
time evolution from the edge of the barrier proceeds into the adjacent
minimum with near certainty, then a trajectory initiated on the
barrier will relax into stable state A with probability
$p_\mathrm{A}=1-y/b$, where $y$ is the initial distance from A and $b$
is the width of the barrier. Similarly, the probability of relaxing
first into state B is $p_\mathrm{B}=y/b$. A standard shooting move
from the barrier region then yields a reactive trajectory with
probability
\begin{equation}
P_\mathrm{acc}=\frac{1}{b}\int_0^b \mathrm{d} y\,p_\mathrm{A}p_\mathrm{B}=\frac{1}{6}\,.
\end{equation}
This value of the acceptance rate should correspond to near optimal
sampling of the transition path ensemble.\cite{Dellago2002} A
problematically low acceptance rate would only arise if one were to 
sample trajectories of insufficient length, i.e., paths shorter
than typical spontaneous transitions.

In our TPS simulations, only momenta (and not particle
positions) are disturbed in the shooting moves, with each
particle's momentum changed in each direction by an amount drawn from
a Gaussian distribution of standard deviation $\Delta p$ (followed by
rescaling of all momenta to enforce energy conservation).\cite{Dellago1998}
We conduct standard shooting moves with values of $\Delta p$ of $10^{-1}$, $10^{-5}$, and $10^{-10}$, as well as
precision shooting moves with $\Delta p$ ranging in size from
$10^{-10}$ to $10^{-300}$. The latter are implemented using helper
displacements  with $\Delta p=10^{-7}$ and are rescaled every time they
reach twice their original size. For the system size studied here, the initial magnitudes of the resulting displacement vectors are larger than the corresponding value of $\Delta p$ by a factor of roughly 34 on average. For instance, $\Delta p=10^{-7}$ results in displacement vectors with an initial size of about $\vert\delta x_0\vert=3.4\times10^{-6}$. For each set of sampling parameters, we attempt 50,000 Monte Carlo moves in trajectory space. Roughly half of these trial moves
are generated by shooting. The other half are generated by a
procedure called ``shifting'',\cite{Dellago2002} in which short trajectory
segments are added to and subtracted from the ends of an existing path.

\begin{figure}
\includegraphics[width=0.4\textwidth]{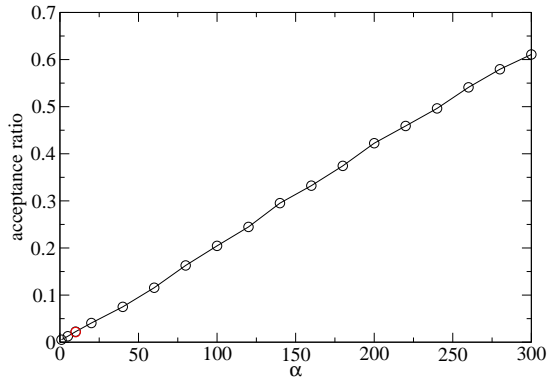}
\caption{\label{acc_fig} 
Fraction of shooting moves for our model isomerization process that
are accepted in long TPS simulations. Acceptance ratios are shown for
shooting displacements of various sizes, $\Delta p=10^{-\alpha}$, implemented using the standard shooting algorithm for values of $\alpha$ of 1, 5, and 10, and the precision shooting algorithm for $\alpha\geq 10$. For $\alpha=10$, the result
obtained from the precision shooting algorithm (red) is effectively indistinguishable from the one obtained with standard shooting.}
\end{figure}
\begin{figure}
\includegraphics[width=0.4\textwidth]{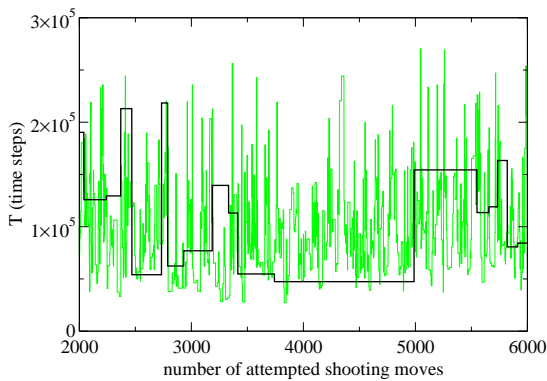}
\caption{\label{samp_fig} 
Variation in isomerization transition time $T$ over the course of long TPS runs.
The bold black line 
shows results for a simulation using shooting displacements with 
$\Delta p=10^{-1}$,
while the thin green line corresponds to
$\Delta p=10^{-100}$.}
\end{figure}

Figure \ref{acc_fig} shows the 
fraction of attempted shooting moves that are accepted in TPS
simulations of the diatomic isomerization with a rough barrier.
While standard shooting moves are accepted with low frequency, any
desired acceptance ratio can be obtained by using the precision
shooting technique. Figure \ref{samp_fig} shows 
changes in transition time $T$ over the course of
two TPS runs with shooting displacements of
$\Delta p=10^{-1}$ and $\Delta p =10^{-100}$. 
A dramatic difference in the efficiency of generating
qualitatively different trajectories for the two cases is evident.

\begin{figure}
\includegraphics[width=0.4\textwidth]{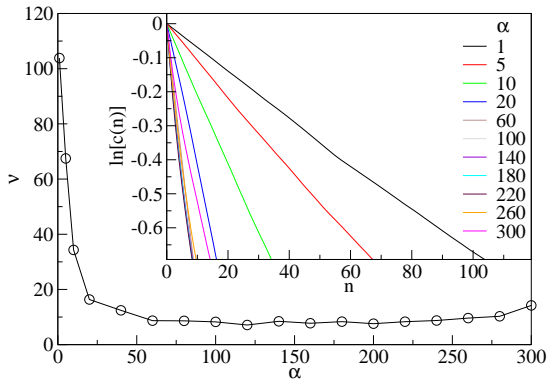}
\caption{\label{ttauto_fig} 
Number of shooting moves required to generate a statistically
independent isomerization trajectory using shooting displacements of
various sizes, with $\Delta p=10^{-\alpha}$. Inset shows decay
of correlation $c(n)$ in the transition time $T$ following $n$
attempted shooting moves. The ``decorrelation time'' $\nu$ is defined
as the value of $n$ beyond which $c(n)<1/2$.}
\end{figure}

To assess the improvement in sampling efficiency achieved with
precision shooting, we quantify the computational effort necessary
to generate statistically independent transition
pathways. More specifically, we
calculate the autocorrelation function
\begin{equation}
c(n) = \frac{\langle \delta T(0)\delta T(n)\rangle}{\langle \delta T^2\rangle}\,,
\end{equation}
where $\delta T(n)=T(n)-\langle T\rangle$ is the deviation of the
transition time after the $n$-th shooting move from its average
$\langle T\rangle$, as calculated from all collected trajectories.\cite{Dellago2002,Bolhuis2003} 
Rapid
decay of $c(n)$
indicates an efficient sampling of trajectories. Figure
\ref{ttauto_fig} shows the logarithm of $c(n)$ for different shooting
displacement
magnitudes
along with the ``decorrelation time'' $\nu$, defined as
the number of successive shooting moves after which the correlation
function
decays to
a value 
less than $1/2$. The maximal sampling efficiency is 
achieved
for shooting displacements with $\Delta p\approx 10^{-100}$.
Improvement over the largest displacement we considered ($\Delta p=10^{-1}$)
is more than ten-fold.
Following Bolhuis\cite{Bolhuis2003}, we also investigate as a measure of
decorrelation changes in the bond length $x$ midway in time through the crossing
event. Decay of correlations in this quantity, and the implied dependence of sampling efficiency on
shooting displacement size, mirror those reported for the transition time $T$.

As Fig.~\ref{ttauto_fig} illustrates, sampling is 
comparably
efficient for a broad range of displacement sizes between $\Delta p=10^{-60}$
and $\Delta p=10^{-260}$. In this regime, the efficiency gain 
due to increased acceptance rates for smaller shooting moves
is compensated 
almost exactly
by the efficiency loss
due to increased similarity between 
the shooting trajectory 
and
its base trajectory. Using equation (\ref{exp_eq}), 
the time $T_\mathrm{id}$ over which a shooting trajectory with
displacement size $10^{-\alpha}$ cannot be resolved from its base
trajectory
can be approximated by the time 
required for the displacement size to reach $10^{-15}$,
\begin{equation}
T_\mathrm{id}\approx \frac{1}{\lambda_1}\ln\frac{10^{-15}}{10^{-\alpha}}= (\alpha-15)\frac{\ln10}{\lambda_1}\,.
\end{equation}
For our system, $1/\lambda_1\approx150\,\Delta t$ and therefore
$T_\mathrm{id}\approx10^5$ time steps for the smallest displacement with
$\Delta p=10^{-300}$. Even for this small displacement size, $T_\mathrm{id}$ is
only 30\% of the total trajectory length $L$ and efficient sampling is
still possible. If $\delta x_0$ is decreased further, $T_\mathrm{id}$
will become comparable to $L$ and sampling efficiency will decrease
accordingly.\footnote{To extend the precision shooting algorithm to
shooting displacements smaller than $10^{-308}$, the smallest
representable number in double precision, exponents can be
conveniently stored separately as integer numbers.} For a displacement
with $\Delta p=10^{-60}$, the largest size that leads to optimum efficiency,
$T_\mathrm{id}$ amounts to only 5\% of the total trajectory length.

\section{\label{sec:meta} Metastable intermediate states}
\begin{figure}
\includegraphics[width=0.4\textwidth]{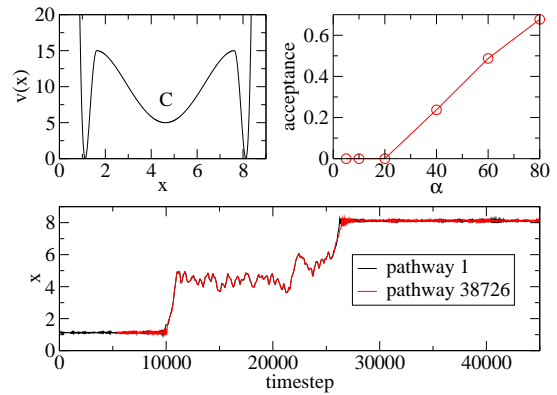}
\caption{\label{inter_fig} TPS simulation of isomerization dynamics
which must proceed through a deep intermediate energy minimum.  Top
left: Interaction potential 
between particles \#1 and \#2. Top right: Acceptance ratio of shooting
moves as a function of displacement size, $\Delta p=10^{-\alpha}$. Bottom: Bond length $x$ as a function of time for the starting
trajectory and for a pathway obtained after many shooting and shifting
moves. The two trajectories have been shifted in time
to highlight their similarity in the vicinity of the intermediate
state.}
\end{figure}
By extending the time span over which a shooting trajectory 
tracks
its base trajectory, the algorithm proposed in this work can
substantially increase the efficiency of TPS simulations that suffer
from poor acceptance of 
shooting moves. 
The method is fully
consistent with deterministic dynamics and faithfully reproduces the
divergent behavior of arbitrarily small displacements in phase space.
We emphasize, however, that the method does not solve all problems
whose primary symptom is a low shooting acceptance rate. Most
importantly, it does not overcome challenges associated with
metastable intermediate states. In this
section we explore the this difficulty in the context of diatomic
isomerization.

In order to explore the consequences of metastable intermediates, we have
modified the diatomic potential $v(x)$ to include a deep minimum
midway between contracted and extended states
(see Fig.~\ref{inter_fig}),
\begin{equation}
v(x)=\left\{ \begin{array}{ll} 	h_1\left[1-q(x)^2/w_1^2\right]^2 & \mathrm{if}\; q(x)<0,\\ 
				h_1\left[1-(q(x)-2w_2)^2/w_1^2\right]^2 & \mathrm{if}\; q(x)>2w_2,\\
				h_1-h_2\left[1-(q(x)-w_2)^2/w_2^2\right]^2 & \mathrm{else}.
\end{array}\right. 
\end{equation}
Here, $q(x)=x-(r_\mathrm{c}+w_1)$, $w_1=0.5$, $w_2=3$, $h_1=15$, and $h_2=10$.
Limited by machine precision, standard shooting moves
fail completely in this case:
Even shooting moves initiated near the intermediate minimum C
rapidly separate from their base trajectories and with high
probability do not escape to stable state A or B.
Only with the precision shooting
technique, using a displacement size smaller than $10^{-20}$, are we
able to conduct successful shooting moves. 
This success does not indicate,
however, that trajectory space is sampled efficiently: A comparison of
the
first
trajectory\footnote{A first trajectory is constructed in the following
way: From the border of state A, trajectories are shot into the
intermediate state, which is divided into small windows along the
direction of the coordinate $x$. Starting with the first of these
windows, trajectories are accepted if they cross the border to the
next one. After accepting a few such trajectories, the simulation
moves on to the next window, eventually leading to a trajectory that
crosses the intermediate from A to B. Note that a similar procedure is
used in forward flux sampling.\cite{Wolde2005}} with a pathway
obtained after many thousands of shooting moves shows that those parts
of the trajectory spent within the intermediate are not resampled at
all (see Fig.~\ref{inter_fig}); they are numerically identical.

Transitions involving strongly metastable intermediates are in fact
fundamentally problematic for TPS methods, unless the dynamics of
intermediates' appearance and disappearance can be identified as
distinct kinetic substeps.
If the typical time spent in C is manageable in a
computer simulation, then the intermediate does not pose a problem
even to the standard shooting move. If, on the other hand, the free
energy barriers 
delimiting the intermediate state are large compared to typical
thermal excitations, then escaping C will itself
be a rare event. In 
such cases, typical transitions from A to B require at least two
unlikely fluctuations (activating entry and exit of each intermediate
state), well separated in time. Any shooting move that perceptibly
modifies dynamics between these rare fluctuations will be rejected
with high probability. Precision shooting can readily generate subtly
modified pathways that remain reactive but cannot be expected to
effectually switch between reactive trajectories that follow
substantially different courses through the intermediate state.
As TPS leaves 
a system's natural dynamics
unchanged, it can eliminate only the largest time scale
associated with a rare event. Without resorting to methods that
prescribe in some sense the detailed route between stable states, one
can overcome the challenge of metastable intermediates with TPS only
by subdividing transition dynamics into several steps, each of which
involves a single dynamical bottleneck.

\begin{acknowledgments}
This work was supported by the Austrian Science Fund (FWF) within the
Science College "Computational Materials Science" under grant W004,
and by the Chemical Sciences, Geosciences, and Biosciences Division of
the U.S. Department of Energy.
\end{acknowledgments}

\end{document}